\pdfoutput=1

\documentclass[10pt,prl,aps,amssymb,amsmath,tightenlines,showpacs,twocolumn]{revtex4}

\usepackage{graphicx}

\newcommand{\cL}{\ensuremath{\mathcal{L}}}
\newcommand{\cP}{\ensuremath{\mathcal{P}}}
\newcommand{\cT}{\ensuremath{\mathcal{T}}}
\newcommand{\cH}{\ensuremath{\mathcal{H}}}
\newcommand{\cPT}{\ensuremath{\mathcal{PT}}}
\newcommand{\half}{\mbox{$\textstyle{\frac{1}{2}}$}}
\begin{document}

\title{$\cPT$-symmetric interpretation of the electromagnetic self-force}

\author{Carl M. Bender$^{1,2}$\email{cmb@wustl.edu} and
Mariagiovanna Gianfreda$^{1,3}$\email{gianfred@le.infn.it}}
\affiliation{$^1$Department of Physics, Washington
University, St. Louis, MO 63130, USA\\
$^2$Department of Mathematical Science, City University London,
Northampton Square, London EC1V 0HB, UK\\
$^3$Institute of Industrial Science, University of Tokyo, Komaba, Meguro,
Tokyo 153-8505, Japan}

\date{\today}

\begin{abstract}
In 1980 Englert examined the classic problem of the electromagnetic self-force
on an oscillating charged particle. His approach, which was based on an earlier
idea of Bateman, was to introduce a charge-conjugate particle and to show that
the two-particle system is Hamiltonian. Unfortunately, Englert's model did not
solve the problem of runaway modes, and the corresponding quantum theory had
ghost states. It is shown here that Englert's Hamiltonian is $\cPT$ symmetric,
and that the problems with his model arise because the $\cPT$ symmetry is broken
at both the classical and quantum level. However, by allowing the charged
particles to interact and by adjusting the coupling parameters to put the model
into an unbroken $\cPT$-symmetric region, one eliminates the classical runaway
modes and obtains a corresponding quantum system that is ghost free. 
\end{abstract}


\maketitle

The techniques of $\cPT$ symmetry have helped to resolve a number of
long-standing theoretical problems, namely, the apparent violation of unitarity
in the Lee model \cite{R1}, the appearance of ghosts in the Pais-Uhlenbeck model
\cite{R2} and in other field-theory models \cite{R3}, and the instability of the
double-scaling limit in $O(N)$ vector models \cite{R4}. In this Letter we apply
these techniques to the famous old problem of runaway modes in classical
electromagnetism.

An oscillating charged particle emits an electromagnetic field, and because the
particle is charged, it interacts with this field. This interaction is called a
{\it self-force}. The classical motion of the oscillating particle is described
by the third-order differential equation \cite{R5}
\begin{equation}
m\ddot{x}+kx=m\tau\dddot{x},
\label{e1}
\end{equation}
where $x(t)$ represents the position of the particle as a function of time. The
three-derivative term is the radiative force obtained from the Abraham-Lorentz
equation. The restoring force constant of the oscillator is $k$ and the mass of
the particle is $m$. For an electron $m=m_e=9\times10^{-31}\,{\rm kg}$. The
parameter $\tau$ can be expressed as $\tau=(4/3)r_q/c$, where $r_q$ is the
classical radius of the charged particle; that is, the radius outside of which
the electric field energy is equal to the rest-mass energy. Thus, $\tau$ is the
time for light to travel across the particle. For an electron $\tau_e=6\times
10^{-24}\,{\rm sec}$.

The solutions to (\ref{e1}) suffer from the physical instabilities of runaway
modes (solutions that grow exponentially with time $t$) and pre-acceleration
\cite{R5}. These behaviors imply that the energy of the particle is not
conserved. Thus, the equation of motion (\ref{e1}) cannot be derived from a
time-independent Hamiltonian. However, Englert \cite{R6} followed the approach
that Bateman used for the damped harmonic oscillator \cite{R7} and constructed a
Hamiltonian system by introducing a time-reversed version of (\ref{e1}):
\begin{equation}
m\ddot{y}+ky=-m\tau\dddot{y}.
\label{e2}
\end{equation}
It is remarkable that even though the $x$ and $y$ equations are separately
nonconservative and are noninteracting, if they are considered together as one
system, then (\ref{e1}) and (\ref{e2}) {\it can} be derived from a Hamiltonian.
Englert interpreted the $y$ particle as an anti-$x$ particle (that is, as a
charge-conjugate version of the $x$ particle). (In their study of the damped
oscillator, Alfinito and Vitiello \cite{R8} also treated their $y$ particle as a
charge-conjugated $x$ particle.) Unfortunately, Englert's construction failed to
solve the problem of runaway solutions to the classical equations. Englert also
studied the quantized version of his classical system and discovered additional
problems with his model, namely, that the energy spectrum is not bounded below
and that there are ghost states (states of negative norm) in the Hilbert space.

We can understand the problems that Englert encountered at the classical level
(runaway modes) and at the quantum level (ghost states) if we re-examine his
work from the point of view of $\cPT$ invariance. By appending (\ref{e2}),
Englert created a system that is $\cPT$ symmetric. However, while a
$\cPT$-symmetric Hamiltonian system has a conserved energy, such a
system can be in one of two possible states: (i) an {\it unbroken} state in
which the classical system is in equilibrium and its frequencies are real
and the corresponding quantum system has real energies and a Hilbert space
with a positive inner product, or (ii) a {\it broken} state in which the
classical system is not in equilibrium because some of its frequencies are
complex and the corresponding quantum system has complex energies and a Hilbert
space with ghosts. The presence of runaway modes in the classical system is a
clear signal that Englert's model is in a broken $\cPT$-symmetric state.

This Letter addresses the problems with Englert's Hamiltonian by allowing the
$x$ and $y$ particles to {\it interact}. A typical $\cPT$-symmetric system can
go from a broken to an unbroken state as the coupling parameters of the system
are varied \cite{R9}. Accordingly, we introduce coupling constants that describe
the interaction between the $x$ and $y$ particles. We then find the region of
the coupling constants in which the system is in an unbroken $\cPT$-symmetric
state. When the system is in equilibrium, there are no runaway solutions to the
equations of motion. We also present substantial evidence that the corresponding
quantum system has a positive real spectrum and a Hilbert space with a positive
inner product. 

In physical terms, a $\cPT$-symmetric system is one for which the loss and gain
are balanced \cite{R10,R11}. Introducing a $\cPT$-symmetric interaction between
the $x$ and $y$ particles solves the problem of runaway modes because as one
particle gains energy from the electromagnetic field, the other particle loses
an equivalent amount of energy. The condition of {\it unbroken} $\cPT$ symmetry
is achieved by coupling the particles sufficiently strongly so that the energy
can flow fast enough from one particle to the other to maintain equilibrium.

In this Letter we construct a classical $\cPT$-symmetric Hamiltonian that
describes a coupled system of $x$ and $y$ particles and we identify the broken
and unbroken regions of $\cPT$ symmetry; that is, the regions for which there
are runaway modes and the regions for which there are no runaway modes. We then
quantize the system. Our analysis of the ground state of the quantum system and
our previous study of coupled damped and undamped $\cPT$-symmetric oscillators
\cite{R9} suggests that because the equations of motion are linear the quantum
system has exactly the same parametric regions of broken and unbroken $\cPT$
symmetry as the corresponding classical system.

\vspace{.1cm}
\noindent{\it Classical model}:
Englert showed that the pair of equations (\ref{e1}) and (\ref{e2}) could be
derived from the Lagrangian
\begin{equation}
L=-\half m\tau(\ddot{y}\dot{x}-\dot{y}\ddot{x})+m\dot{x}\dot{y}-kxy.
\label{e3}
\end{equation}
The equations of motion (\ref{e1})-(\ref{e2}) arise from \cite{R12}
\begin{eqnarray}
0&=&\frac{\delta L}{\delta x}=\frac{\partial L}{\partial x}-\frac{d}{dt}\frac{
\partial L}{\partial\dot{x}}+\frac{d^2}{dt^2}\frac{\partial L}{\partial
\ddot{x}},\nonumber \\
0&=&\frac{\delta L}{\delta y}=\frac{\partial L}{\partial y}-\frac{d}{dt}\frac{
\partial L}{\partial\dot{y}}+\frac{d^2}{dt^2}\frac{\partial L}{\partial
\ddot{y}}.
\label{e4}
\end{eqnarray}
We construct a Hamiltonian from the Lagrangian (\ref{e3}) by using the formula
$H=\sum_a \dot{a}p_a-L$, where $a=x,\dot{x},y,\dot{y}$:
\begin{equation}
H=\frac{ps-rq}{m\tau}+\frac{2rs}{m\tau^2}+\frac{pz+qw}{2}-\frac{mzw}{2}+kxy,
\label{e5}
\end{equation}
where we have introduced the variables $z=\dot{x}$, $w=\dot{y}$, $p=p_x$, $q=
p_y$, $r=p_z$, and $s=p_w$. In this model the two systems $x(t)$ and $y(t)$ are
noninteracting, as we can see from (\ref{e1}) and (\ref{e2}). This system is
$\cPT$ symmetric, where the effects of parity and time reversal are given in
Table \ref{T1}. However, it is not invariant under parity reflection $\cP$ or
under time reversal $\cT$ alone.

\begin{table}[h!]
\begin{center}
\renewcommand{\arraystretch}{1.5}
\begin{tabular}{c|c|c|c|c|c|c|c|c| }
& \,\,\,\,x\,\,\,\, &\,\, y\,\, \, &\,\, z\,\, \,&\,\, w\,\,\,&\,\,\,
p\,\,\,\,&\,\,\,\,q\,\,\,\,\,&\,\,\, r\,\,\,\, &\,\,\,\,s\,\,\,\,\\ \hline
$\cP$ & y & x & w & z & q & p & s & r \\
$\cT$ & x & y & -z & -w & -p & -q & r & s \\
$\cPT$ & y & x & -w & -z & -q & -p & s & r
\end{tabular}
\end{center}
\caption{Behaviors of the variables $x$, $y$, $z$, $w$, $p$, $q$, $r$, and $s$
in the Hamiltonian (\ref{e5}) under space reflection $\cP$, time reversal $\cT$,
and combined $\cPT$.}
\label{T1}
\end{table}

\begin{figure}[t!]
\includegraphics[scale=0.4]{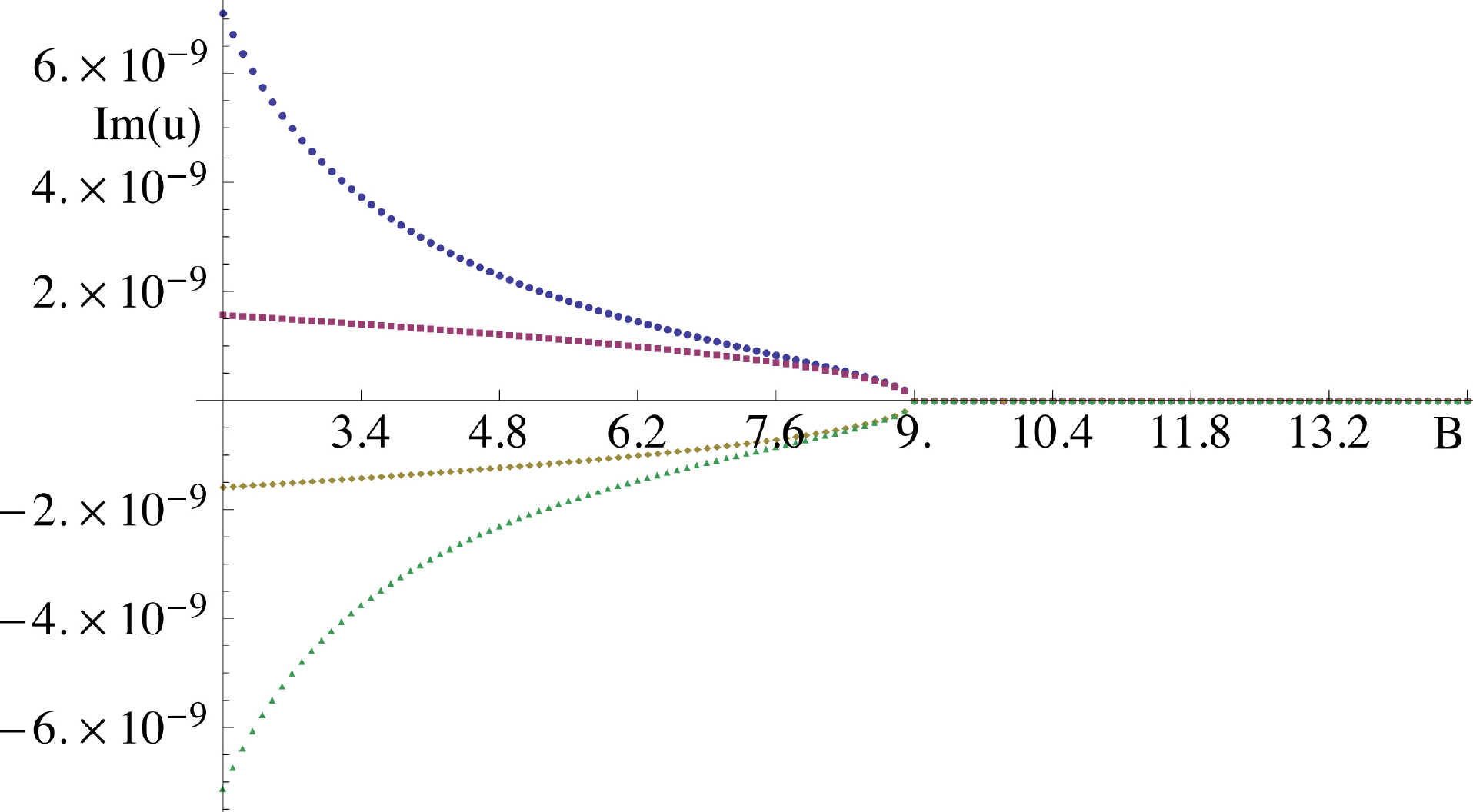}
\caption{Imaginary part of the roots $\lambda$ of the classical polynomial
(\ref{e8}) and imaginary part of the roots $u$ of the quantum polynomial
(\ref{e24}) for the case of a physical electron $m=m_e$ and $\tau=\tau_e$. We
choose $A=1.1$. The parameter $B$ is scaled by a factor $10^{-31}$. We can see
that the $\cPT$ transition arises at exactly the same value of the coupling
parameter $B$ for both the classical and the quantum model; this value is
approximately equal to the mass of the electron $m_e$.}
\label{F2}
\end{figure}

Following the approach used in Ref.~\cite{R9}, we introduce two interaction
terms in the equations of motion (\ref{e1})-(\ref{e2}):
\begin{eqnarray}
m\tau\dddot{x}-m\ddot{x}-kx&=&Ay+B\ddot{y},\nonumber\\
m\tau\dddot{y}+m\ddot{y}+ky&=&-Ax-B\ddot{x}.
\label{e6}
\end{eqnarray}
(There are in fact 14 possible quadratic interaction terms that we could
introduce in the Hamiltonian that governs the theory and which do not increase
the order of the equations of motion. The equations above are the simplest in
which we observe a transition between regions of broken and unbroken $\cPT$
symmetry.) Figure~\ref{F2} shows that at the transition from broken to unbroken
$\cPT$ symmetry the size of the coupling parameter $A$ is of order $k$ and the
coupling parameter $B$ is of order $m$.

There are two conserved quantities for the system (\ref{e6}). The first quantity
$E_1$ is obtained by multiplying the first and the second equation of (\ref{e6})
by $\dddot{y}$ and $\dddot{x}$:
\begin{eqnarray}
E_1&=&m\ddot{x}\ddot{y}+k(x\ddot{y}+\ddot{x}y-\dot{x}\dot{y})\nonumber\\
&&\quad+\frac{A}{2}(2x\ddot{x}+2y\ddot{y}-\dot{x}^2-\dot{y}^2)
+\frac{B}{2}(\ddot{x}^2+\ddot{y}^2).\nonumber
\end{eqnarray}
The second quantity $E_2$ is obtained by multiplying the first and the second
equation of (\ref{e6}) by $\dot{y}$ and $\dot{x}$:
\begin{equation}
E_2=m\tau(\ddot{x}\dot{y}-\dot{x}\ddot{y})+m\dot{x}\dot{y}+kxy
+\frac{A}{2}(x^2+y^2)+\frac{B}{2}(\dot{x}^2+\dot{y}^2).\nonumber
\end{equation}
The Lagrangian for the system (\ref{e6}) is $\cL=L-\half A(x^2+y^2)+\half B
(\dot{x}^2+\dot{y}^2)$ and the Hamiltonian is given by
\begin{equation}
\cH=H+\frac{A}{2}(x^2+y^2)+\frac{B}{m\tau}(rw-sz).
\label{e7}
\end{equation}

\begin{figure}[h!]
\includegraphics[scale=0.4]{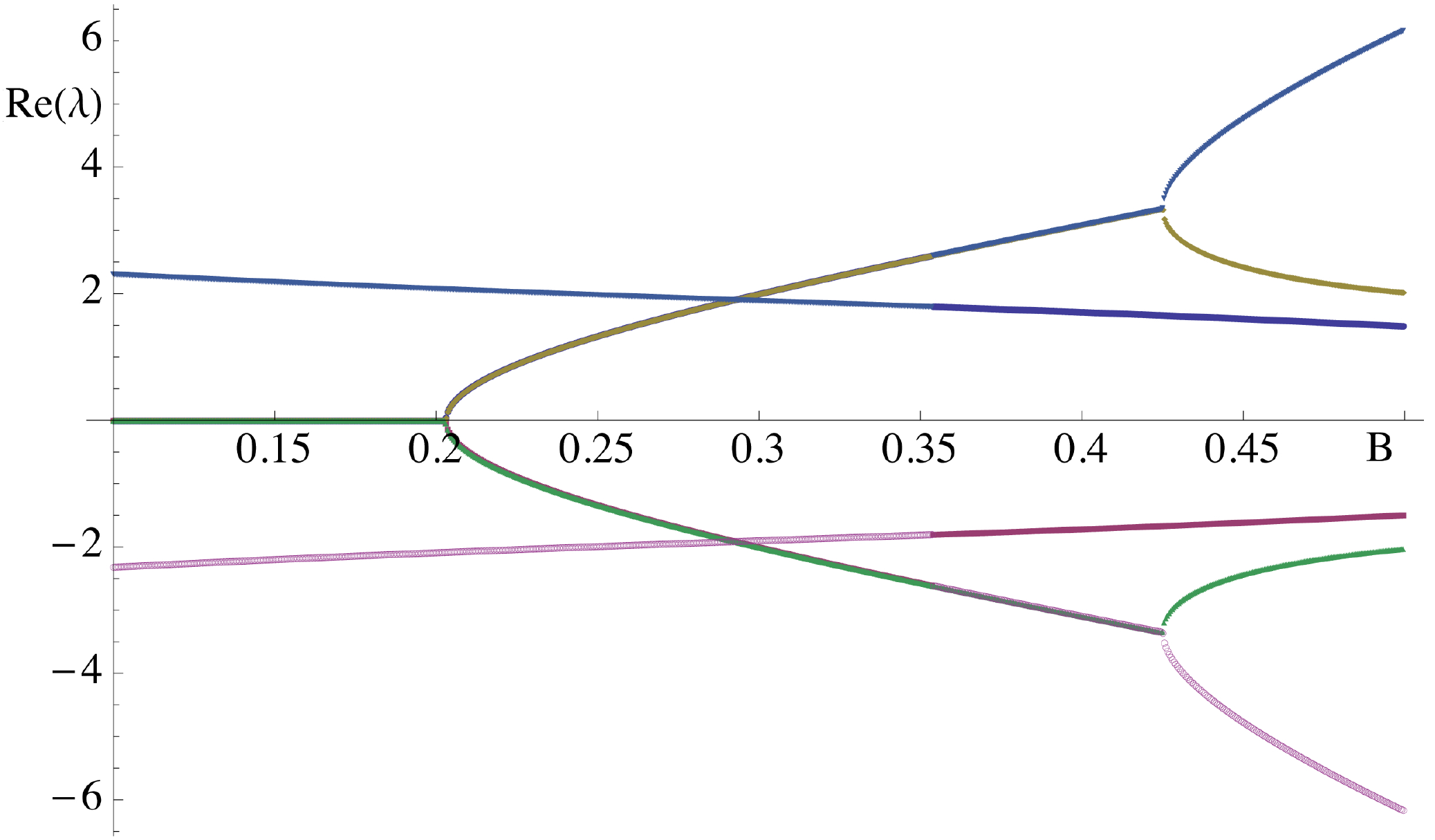}
\caption{Imaginary parts (upper panel) and real parts (lower panel) of the six
roots $\lambda$ of the polynomial (\ref{e8}) for $m=0.3$, $\tau=0.2$, $A=1.5$
plotted as a function of $B$. There is a transition from broken to unbroken
$\cPT$ symmetry as $B$ increases from $B<0.425$ to $B>0.425$. The roots are real
when $B>0.425$.}
\label{F1}
\end{figure}

To determine whether the system (\ref{e6}) is in a broken or an unbroken
$\cPT$-symmetric phase, we seek solutions of the form $x(t)=\alpha\,e^{i\lambda
t}$ and $y(t)=\beta\,e^{i\lambda t}$ ($\alpha,\,\beta$ constants). The frequency
$\lambda$ satisfies the sixth-degree
polynomial equation
\begin{equation}
0=\lambda^6+\frac{m^2-B^2}{m^2\tau^2}\lambda^4+\frac{2AB-2m}{m^2\tau^2}
\lambda^2+\frac{1-A^2}{m^2\tau^2}.
\label{e8}
\end{equation}
This secular equation has real coefficients because the Hamiltonian (\ref{e7})
is $\cPT$ symmetric. The $\cPT$ symmetry is unbroken if we can find parameters
$(A,B)$ for which all six roots of (\ref{e8}) are real (so there are no
exponentially growing modes). We let $k=1$ without loss of generality. Then if
we take $m=0.3$, $\tau=0.2$, and $A=1.5$, we can see from Fig.~\ref{F1} that
the region of unbroken symmetry is $B>0.425$, the region of broken $\cPT$
symmetry is $B<0.425$, and the $\cPT$ transition occurs at $B=0.425$.

\begin{figure}[h!]
\includegraphics[scale=0.4]{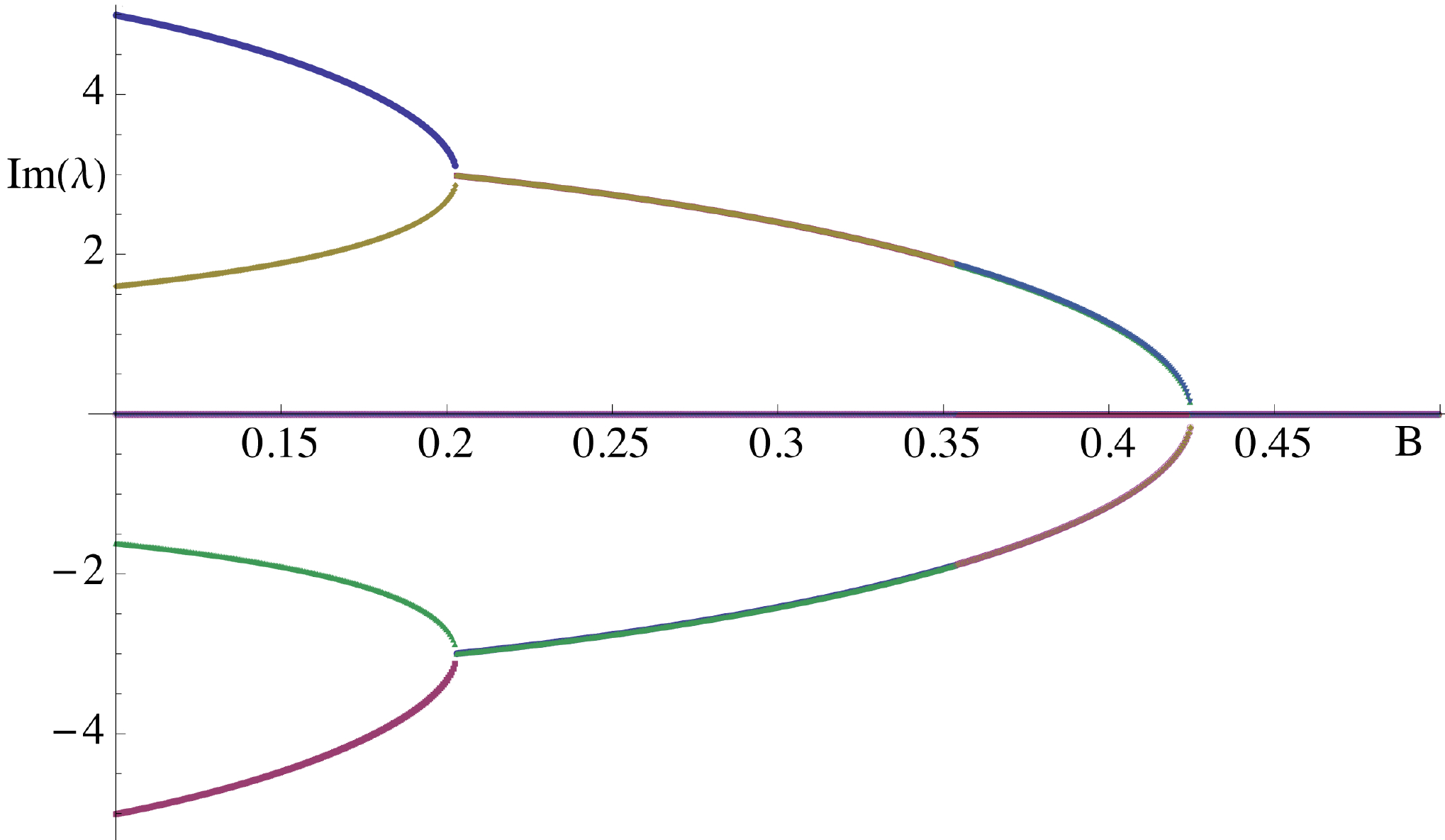}
\end{figure}
\vspace{-0.5cm}

\vspace{.1cm}
\noindent{\it Quantization of the model}: The states of the quantized version of
the Hamiltonian $\cH$ in (\ref{e7}) satisfy the time independent Schr\"odinger
equation
\begin{equation}
\hat{\cH}\psi_0=E\psi_0,
\label{e9}
\end{equation}
where the coordinate-space Hamiltonian is
\begin{eqnarray}
\hat{\cH}&=&\frac{iB(z\partial_w-w\partial_z)}{m\tau}-\frac{2\partial^2_{zw}}{m
\tau^2}+\frac{\partial^2_{yz}-\partial^2_{xw}}{m\tau}-\frac{mzw}{2}\nonumber\\
&& \quad-\frac{iw\partial_y+iz\partial_x}{2}+kxy
+\frac{Ax^2}{2}+\frac{Ay^2}{2}.
\label{e10}
\end{eqnarray}

For simplicity we restrict our attention to the ground state $\psi_0$ only. In
coordinate space this state has the form of a Gaussian in the four variables
$x$, $y$, $z$, and $w$:
\begin{eqnarray}
\psi_0&=&\exp\Big\{-\frac{m}{2}\Big[\frac{a+ib}{2\tau}x^2+\frac{a-ib}{2\tau}
y^2+\frac{\tau(c+id)}{2}z^2\nonumber\\ 
&&\!\!\!\!\!\!\!\!\!+\frac{\tau(c-id)}{2}w^2+(u+iv)xz+(iv-u)yw
+\frac{exy}{\tau}\nonumber\\
&&\!\!\!\!\!\!\!\!\!+(g+ih)xw+(ih-g)yz+n\tau zw\Big]\Big\}.
\label{e11}
\end{eqnarray}
If we substitute (\ref{e11}) into (\ref{e9}), we obtain a system of ten coupled
nonlinear algebraic equations for the real coefficients $a$, $b$, $c$, $d$, $e$,
$g$, $h$, $n$, $u$, and $v$:
\begin{eqnarray}
0&=&2(vh-ug)+bh+eu-ag+AQ,\label{e12}\\
0&=&u^2+v^2+g^2+h^2+au+bv-eg+Q,\label{e13}\\
0&=&c^2+d^2+n^2+2Rd+cu+dv\nonumber\\ &&\quad +ng+h+1,\label{e14}\\
0&=&2(cg+nu+ug-dh-Rh)+an+b-ce,\label{e15}\\
0&=&u^2+v^2+g^2-h^2+2(Rv+cu+dv+ng)\nonumber\\ &&\quad+ac+bd-en,\label{e16}\\
0&=&2cn+cg+dh+nu+v,\label{e17}\\
0&=&2(uh+vg)+ah+bg-ev,\label{e18}\\ 
0&=&2(ch+dg+Rg+nv+vg)-a-de+bn,\label{e19}\\
0&=&2(cv-du-Ru+nh+gh)+bc-ad-e,\label{e20}\\
0&=&2(dn+Rn)+dg-ch+nv-u,\label{e21}
\end{eqnarray}
where $Q=2\tau^2/m$, $R=B/m$, and the ground-state energy $E=E_0$ is
$E_0=(n+g)/\tau$.

This system of equations is not easy to solve, but we have devised the following
procedure to do so: First, we introduce new parameters $X$ and $Y$:
\begin{equation}
A=(X+1/X)/2,\quad R=(Y+1/Y)/2.
\label{e22}
\end{equation}
Next, we eliminate the five variables $c$, $n$, $g$, $h$, and $v$ by
substituting
\begin{eqnarray}
c&=&\frac{Y^2+2dY+1}{Y^2-1},\quad n=-\frac{dY^2+2Y+d}{Y^2-1},\nonumber\\
g&=&-\frac{u(X+Y)}{XY+1},\quad h=\frac{u(XY-1)}{X Y+1},\nonumber\\
v&=&\frac{u(Y-X)}{XY+1}
\label{e23}
\end{eqnarray}
into all ten equations. We then observe that (\ref{e17}) and (\ref{e21}) reduce
to $0=0$. Next, we solve (\ref{e15}) and (\ref{e16}) for $a$ and $b$ and we
observe that when these variables are eliminated from the remaining equations,
(\ref{e19}) and (\ref{e20}) become $0=0$. We then solve (\ref{e18}) for $e$ and
eliminate this variable from the other equations. 

At this point, only three equations, (\ref{e12}), (\ref{e13}), and (\ref{e14}),
remain unsolved. We solve (\ref{e14}) for $d^2$ in terms of $d$ and $u$. When we
use this result to simplify the other two equations, we see that (\ref{e12}) and
(\ref{e13}) are redundant. Finally, we solve (\ref{e12}) for $d$ and eliminate
$d$ from (\ref{e14}). This gives a surprisingly simple fourth-degree polynomial
equation for $U=u/(XY+1)$:
\begin{eqnarray}
0&=&\left[16X^2Y(X-Y)(XY-1)(Y^2+1)^2\right.\nonumber\\
&&\,\,\left.+8QXY^2(X+Y)^3(XY+1)\right]U^4\nonumber\\
&&\,\,+\,8QXY(X+Y)(2X^2Y^3-XY^4+X-2Y)U^3\nonumber\\
&&\,\,+\,2QX (Y^4+6Y^2+1)(X-Y)(XY-1)U^2\nonumber\\
&&\,\,-\,2Q^2(X^2-1)Y^2(XY+1)U\nonumber\\
&&\,\,-\,Q^2Y(X-Y)(XY-1).
\label{e24}
\end{eqnarray}

Finally, we obtain the ground-state energy using $E_0=(n+g)/\tau$. Note that
there are two values of $X$ and two values of $Y$ for each value of $A$ and $R$.
Therefore, there are actually four sets of solutions to (\ref{e12})-(\ref{e21}).
Furthermore, since $U$ satisfies a fourth-degree polynomial equation, there are
a total of 16 possible values for the ground-state energy $E_0$. All 16 of
these values are {\it real} when $u$ is real, and this defines the {\it
unbroken} region of $\cPT$ symmetry. In Fig.~\ref{F2} we plot the imaginary
part of the classical frequency $\lambda$ and the imaginary part of $u$
obtained from (\ref{e8}) and (\ref{e24}) for physically realistic values of
the couplings $A$ and $B$. Note that the $\cPT$ transition for the classical
theory and for the quantum theory coincide.

\begin{figure}[t!]
\includegraphics[scale=0.4]{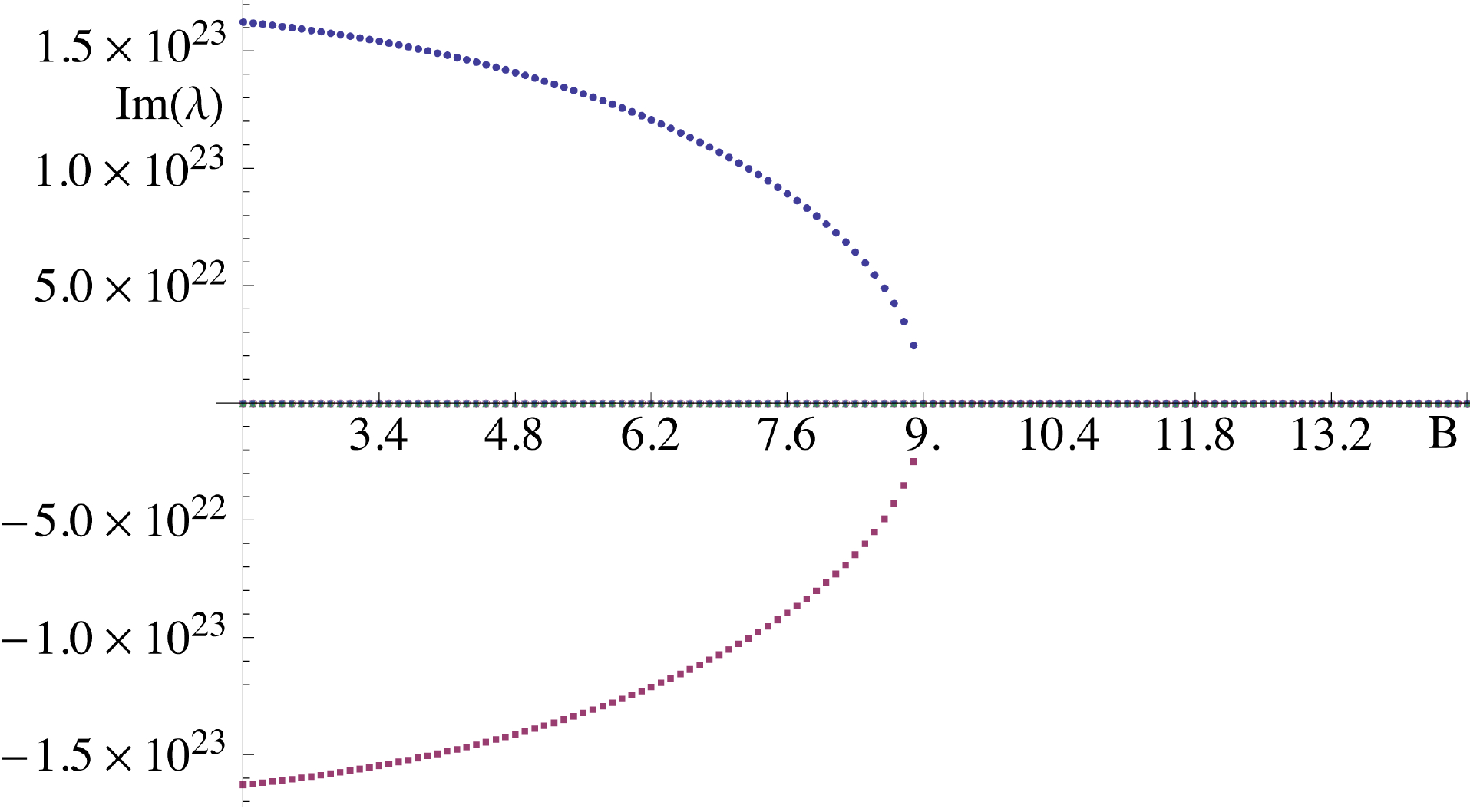}
\end{figure}

While there are 16 possible values for $E_0$, only one value is physically
acceptable. To show this we must calculate the energies of the higher excited
states. The eigenfunctions of the $n$th excited states of $\hat{\cH}$ have
the form of $\psi_0$ in (\ref{e11}) multiplied by a polynomial of degree
$n$ in the variables $x$, $y$, $z$, and $w$. We find that there is a {\it
unique} value of $E_0>0$ for which the entire spectrum of $\hat{\cH}$ is
bounded below by $E_0$. The details of this calculation are presented in a
longer and more detailed paper \cite{R13}. Ref.~\cite{R13} also describes the
Stokes wedges in the complex-$x$, $y$, $z$, and $w$ planes inside of which the
eigenfunctions are normalizable and it demonstrates the orthogonality of 
the eigenfunctions.

We conclude with some conjectural remarks. The conventional way to explain
why the classical system (\ref{e1}) can have runaway modes, which
appear to violate the conservation of energy, is to argue that there is an
infinite source of energy in the electromagnetic field of the electron. (This is
because an infinite amount of work is required to assemble a pointlike electron
by bringing in charge from infinity, and this work is stored in the
electromagnetic field.) There are no runaway modes in the $\cPT$-symmetric
system (\ref{e6}) in the unbroken region. The underlying reason that the energy
is unavailable to support the instability associated with runaway modes is that
there are interference effects
between the strongly coupled $x$ and $y$ particles in the unbroken
$\cPT$-symmetric region. We conjecture that one can characterize the difference
between the broken and unbroken regions of $\cPT$-symmetric systems by saying
that the classical charge renormalization is correspondingly infinite or finite.
Perhaps this distinction applies to $\cPT$-symmetric quantum systems as well.

CMB thanks the U.S.~Department of Energy and MG thanks the Fondazione Angelo
Della Riccia for financial support.

\end{document}